# Three-dimensional charge density wave and robust zero-bias conductance peak inside the superconducting vortex core of a kagome superconductor $CsV_3Sb_5$

Zuowei Liang[1#], Xingyuan Hou[2#], Fan Zhang[3,2#], Wanru Ma[1], Ping Wu[1], Zongyuan Zhang[2], Fanghang Yu[1], J.–J. Ying[1], Kun Jiang[3], Lei Shan[2,4*], Zhenyu Wang[1*] and X.-H. Chen[1,5,6*]

[1]*Department of Physics and Chinese Academy of Sciences Key laboratory of Strongly-coupled Quantum Matter Physics, University of Science and Technology of China, Hefei, Anhui 230026, China*
[2]*Information Materials and Intelligent Sensing Laboratory of Anhui Province, Institutes of Physical Science and Information Technology, Anhui University, Hefei 230601, China*
[3]*Beijing National Laboratory for Condensed Matter Physics, and Institute of Physics, Chinese Academy of Sciences, Beijing 100190, China*
[4]*Key Laboratory of Structure and Functional Regulation of Hybrid Materials of Ministry of Education, Anhui University, Hefei 230601, China*
[5]*CAS Center for Excellence in Quantum Information and Quantum Physics, Hefei, Anhui 230026, China*
[6]*Collaborative Innovation Center of Advanced Microstructures, Nanjing 210093, China*

*Correspondence and requests for materials should be addressed to Z.W. (zywang2@ustc.edu.cn), L. S. (lshan@ahu.edu.cn) or X.-H.C. (chenxh@ustc.edu.cn).

The transition-metal-based kagome metals provide a versatile platform for correlated topological phases hosting various electronic instabilities. While superconductivity is rare in layered kagome compounds, its interplay with nontrivial topology could offer an engaging space to realize exotic excitations of quasiparticles. Here, we use scanning tunneling microscopy (STM) to study a newly discovered $Z_2$ topological kagome metal $CsV_3Sb_5$ with a superconducting ground state. We observe charge modulation associated with the opening of an energy gap near the Fermi level. When across single-unit-cell surface step edges, the intensity of this charge modulation exhibits a $\pi$-phase shift, suggesting a three-dimensional 2×2×2 charge density wave ordering. Interestingly, a robust zero-bias conductance peak is observed inside the superconducting vortex core on the Cs 2×2 surfaces that does not split in a large distance when moving away from the vortex center, resembling the Majorana bound states arising from the superconducting Dirac surface states in $Bi_2Te_3/NbSe_2$ heterostructures. Our findings establish $CsV_3Sb_5$ as a promising candidate for realizing exotic excitations at the confluence of nontrivial lattice geometry, topology and multiple electronic orders.

The kagome nets consisting of 3d transition metal ions are emerging as a new frontier for exploring novel correlated and topological electronic states[1-10]. Thanks to its special lattice geometry, a kagome lattice naturally possesses Dirac cones and flat band with quenched kinetic energy[11-12], leading to interaction-driven topological many-body phenomena. Experimentally, 3D magnetic Weyl semimetal[4, 13-15] and 2D Chern insulator[9] phases have been realized in kagome lattices compounded with a net magnetization. On the other hand, depending on the electron filling level, various electronic states ranging from density waves to superconductivity have also been theoretically predicted[11, 16-18]. However, the search for layered kagome compound with a superconducting ground state remains a great challenge, leaving much of the related field unexplored.

Recently, a new family of layered kagome metals $AV_3Sb_5$ (A= K, Rb, Cs) has been discovered with a superconducting ground state[19-22]. The normal state of these compounds has been theoretically identified to host a $Z_2$ topological invariant, with topological Dirac surface states near the Fermi energy (Dirac energy $E_D$ ~40meV) and the calculated band structure is supported by the angle-resolved photoemission measurements[20]. In addition, these materials all exhibit a clear transport and magnetic anomaly at $T^*$ ~ 78-104K due to the formation of charge order[19-21], suggesting that the electronic interactions could be at play in these compounds. This charge order has been revealed to be unconventional with a chiral anisotropy[23], giving rise to the observed anomalous Hall effect in the absence of long-range magnetic order[24, 25]. When these $Z_2$ topological metals fall into the superconducting state, the natively proximitized superconducting topological surface states may provide a promising route to realize zero energy modes related to Majorana bound states (MBSs)[20, 26], which are the key ingredients for topological quantum computation[27]. Furthermore, signatures of spin-triplet pairing and an edge supercurrent in $Nb/K_{1-x}V_3Sb_5$ devices have been reported[28], suggesting a potential exotic superconducting state in these devices. Therefore, a spectroscopic probe of the electronic orders in $AV_3Sb_5$, particularly for the superconducting state, is highly desired.

Scanning tunneling microscopy (STM), owing to its high spatial topographic resolution and high energy spectroscopic resolution, is an ideal probe to investigate the real space charge ordering and novel quasiparticle excitations in the superconducting state. In this work, we use low-temperature STM to study the kagome superconductor $CsV_3Sb_5$ with the highest $T_C$ ~ 3K in the $AV_3Sb_5$ family. We pinpoint a three-dimensional 2×2×2 charge density wave (CDW) order in $CsV_3Sb_5$, with the opening of an energy gap of ~±20meV at the Fermi level. Inside the superconducting vortex core on the ordered Cs surfaces, we observe a robust, non-split zero-bias conductance peak (ZBCP) within a large distance away from the core center, whose full width at half maximum (FWHM) almost does not change. This ZBCP bears a remarkable resemblance to the MBSs that have been observed in the $Bi_2Te_3/NbSe_2$ heterostructures[29].

Single crystals of $CsV_3Sb_5$ were grown via a self-flux growth method as described in earlier studies[19-20]. The crystals are first characterized by the x-ray diffraction (supplementary information note I; SI 1) and resistivity measurements. As shown in Fig. 1B, a resistivity anomaly appears at 94K, similar to the previous report[20]. The onset of the superconductivity in our samples occurs around 3.5K and zero resistivity can be achieved around 2.7K. The crystals were cold-cleaved *in-situ* at 30 or 80K, and then immediately transferred into the STM head. Chemically etched tungsten tips were used in all the measurements, after being checked on a clean Au (111) surface. STM measurements for the normal state and superconducting state are carried out at 5K and 0.4K, respectively. Spectroscopic data were acquired by the standard lock-in technique at a frequency of 987.5Hz.

$CsV_3Sb_5$ has a layered structure with the space group of P6/mmm and hexagonal lattice constant a=5.5 Å and c=9.3 Å. It consists of a kagome vanadium net interpenetrated by a hexagonal net of Sb1 in-between two honeycomb-structured Sb2 layers, and all sandwiched in between two hexagonal Cs layers (Fig. 1A).

Considering the chemical bonding between V, Sb2 and Cs, the main cleavage plane is in-between the Cs and Sb2 layers, generating either Cs or Sb terminated surfaces. Atomically resolved STM topographies are shown in Fig. 1C and D, with hexagonal and honeycomb lattice respectively. In SI 2, we show a joint area between these two types of lattices, with an atomic step edge where the upper regime shows hexagonal lattice while the lower shows honeycomb lattice. On the basis of the crystalline structure, we can identify them as Cs (Fig. 1C) and Sb (Fig. 1D) surfaces, respectively. One can see tridirectional (3Q) 2*2 superstructure in the real space topographies for both types of surfaces, with additional peaks at half of the Bragg reciprocal lattice vectors in the Fourier transform, similar to the case of $KV_3Sb_5$ (ref. 23). The observed modulation on both Cs and Sb surfaces suggests a 2×2 superlattice pattern in the underlying kagome V plane, as illustrated in Fig. 1A, since the 3d orbitals of V have a large contribution to the total electron density of states (DOS) at low energyies[20, 23]. The cleavage can also break the Cs atomic layer and result in a third type of surface with half amount of the Cs atoms. In fact, this is the most commonly observed surface in our experiments. Fig. 1E shows one of such surfaces, where half of the Cs-chains is alternately missing, as illustrated by the gray lines in Fig. 1a. Interestingly, we can still see clear charge modulation on all the remnant Cs chains (Fig. 1E; more bias-dependent topographic images of half-Cs surface can be found in SI 3). We would like to note that a unidirectional modulation with $4a_0$ or $5a_0$ lattice constant (1Q-charge order; ref. 30), in addition to the 3Q-2*2 order, can be observed on largely-exposed Sb surfaces with only a few Cs atoms on the top (Fig. 1F). Thus, these two types of 3Q- and 1Q- charge orders may save similar total energy and can be influenced by strain [30, 31] or doping, especially near the surface (SI 4).

Typically, in an electronic ordering phase, such as the classic Peierls-CDW state, one could observe a suppression of DOS near the Fermi level. We characterize the DOS by measuring the differential conductance (*dI/dV*) and show the typical spectra far away from defects in Fig. 1 G-J. There are several intriguing features in these spectra. Firstly, although a suppression of DOS near the $E_F$ can be observed for all these surfaces, this gap-like feature is clearer on the Sb surface, and we can deduce an energy scale of ~ ±20meV from the change of slope in the spectral line-shape. On the half-Cs surfaces, additional states appear at low energy (around -12meV), which exhibit spatial variations with the superlattice (SI 5) and change the apparent spectral line-shape. Secondly, the suppression of DOS disappears above 94K as shown in SI 6, suggesting its close relationship with the formation of the superlattice. Thirdly, the additional modulation of the superlattice also leads to pronounced signatures in the spectroscopic images that reverses its intensity across the Fermi energy (SI 5), consistent with the charge density wave (CDW) contribution to STM images observed in many CDW materials[33-35]. Thus, the observed superlattice modulation, the suppression of DOS near $E_F$, and the contrast inversion collectively support a CDW state in $CsV_3Sb_5$. In addition, we find that the minimum of the DOS is located slightly above $E_F$, similar to that observed in $NbSe_2$ (ref. 30). Although the driving mechanism of the CDW order remains unclear in $CsV_3Sb_5$, the particle-hole asymmetric gap indicates that it could not be simply weak-coupled Fermi surface nesting.

With the formation of 2×2 superlattice in the Cs layer, one would expect to see two types of topographies on the half-Cs surface, one with alternating bright and dark atoms in a chain while the other only containing dark Cs atoms without contrast changing (see Fig. 1a for an illustration where the dark-atom chains have been marked with gray lines), and the ratio between these two should be roughly 1:1. In spite of intensive search, only the former case has been observed on our cleaved samples. One possible reason for this deviation could be that the unit-cell of $CsV_3Sb_5$ doubles along the c axis in the CDW state, so that the Cs atoms on the topmost surface reflect the CDW pattern of the underlying V lattice which hosts a π-phase jump. A single-unit-cell surface step could offer a direct probe to check this picture. Figure 2A shows a topographic image including a single-unit-cell surface step edge (with the height of about 9 Å, Fig. 2B), which allows us to simultaneously view the CDW patterns of the upper and lower half-Cs surfaces side-by-side. A close-up view reveals that the CDW pattern exhibits a relative phase shift across the single-unit-cell step edge (Fig. 2C and D; the topographic bright spots correspond to the bright sites in the CDW patterns). To precisely determine the magnitude of the shift, a displacement analysis in reciprocal space is carried out (SI 7), which indicates that the phase shift is actually π. Measurements taken near other two single-unit-cell steps are shown in SI 7, and a similar π-phase jump has been observed. These observations further establish that the charge order in $CsV_3Sb_5$ is a three-dimensional 2×2×2 CDW state. The doubling of unit-cell along the *c* direction may be related to the observed two-dome superconducting phase diagram under pressure[36].

We have performed spectroscopic imaging on the half-Cs surfaces. The Fourier transforms (FTs) of the differential conductance maps reveal clear quasiparticle interference (QPI) patterns (Fig. 2E and F; more data can be found in SI 8). The white circle in the FTs denotes the 3Q-CDW wave vector. The QPI data show two ring-like signals centered at $\Gamma$ point. We find that the outer ring ($q_0$) is almost non-dispersing, which may potentially originate from Friedel oscillations ($2k_F$) or STM set-point effect[37]. The inner one ($q_1$) arises from the intra-pocket scattering of the electron band at $\Gamma$, which has been assigned to the $p_z$ orbital of Sb. The extracted dispersion of $q_1$, together with a parabolic fit, are displayed in Fig.2G. The band bottom is estimated to be -0.6eV from the fitting, consistent with the reported ARPES data[20].

We now investigate the superconducting state of the $CsV_3Sb_5$. At 0.4K, we observe a suppression of DOS at $E_F$ with symmetric coherence peaks forming at ± 0.55meV in the tunneling spectra (Fig. 3A), which we identify to be the superconducting gap. This gap is gradually suppressed at elevated temperatures and eventually disappears at the bulk Tc~ 3K. Interestingly, we find that the superconducting spectra obtained on different types of surfaces are slightly different. As summarized in Fig. 3B, While the spectra acquired on the Cs 2×2 and Sb surfaces have more similar line-shapes, the superconducting spectrum of half-Cs surface shows clearly suppressed superconducting coherence peaks, indicating a larger scattering effect[38]. The overall evolution of the superconducting spectra has been checked on two samples. To gain a quantitative understanding on the superconducting order parameter, we fit the tunneling spectrum of Cs surface with the Dynes model[38], using isotropic s-wave gap, anisotropic s-wave and two isotropic s-wave gap functions, respectively. The results can be found in SI 9.

Next, we check the possible excitations inside a superconducting vortex. The vortices are directly imaged by mapping the conductance inside the superconducting gap with a magnetic field of 0.035T perpendicular to the *ab* plane. Based on the single magnetic flux quanta 2.07 $\times 10^{-15}$ Wb, one should see roughly one vortex in a 230nm ×230nm field of view. Fig.4a shows such a map at zero energy with a single vortex on the Cs surface. We observe a clear ZBCP at the vortex core (Fig. 4B). Differential conductance spectra acquired along two distinct paths (line 1 and line 2, denoted by the arrows in Fig. 4A) across the vortex reveal an intriguing evolution, which are shown in the false-color images of Fig. 4C and D, respectively. When tracking the splitting peaks near the vortex core edge, we find that the start point of the splitting is located ~ 35nm away from the vortex core (close-up of the splitting spectra is shown in SI 10), and the FWHM of the ZBCP bears little change in this range within our STM resolution ~0.23meV (SI 11). This observation is in sharp contrast to the vortex core states in conventional s-wave superconductors, such as $NbSe_2$ (ref. 39), where an "X"-type splitting has been widely observed, but is similar to the "Y"-type splitting observed in 4-6 QL $Bi_2Te_3/NbSe_2$ heterostructures (ref. 29; a direct comparison has been shown in SI 10). We would like to emphasize that the ZBCP inside the vortex core is not induced by disorder for two reasons. First, at zero field, homogenous superconducting spectra can be observed along the same trace of line 1 (SI 12), suggesting that the ZBCP indeed represents quasiparticle excitations inside a vortex. In fact, we never see in-gap state that extends hundred nanometers at 0 T in our samples. Second, this "Y"-type splitting of ZBCP is repeatable once the vortex is located on the Cs surfaces (SI 13).

As we have mentioned above, the largely-explored Sb surfaces exhibit additional 1Q modulation, which is in contrast to the Cs surfaces that show only 3Q-2×2 CDW. Interestingly, a distinct evolution of the vortex-core state is observed on the Sb surface. As shown in Fig. 4G and H, the ZBCP inside the vortex core obtained on the Sb surface splits faster when moving away from the core center, exhibiting a clear "X"-type evolution that is more similar to the conventional case. As a result, the apparent width of the vortex-core state also shows remarkable spatial dependence, in contrast to that observed on the Cs surface (Fig. 4F). For a better comparison, the spatial evolutions of vortex-core states observed on the Cs surface, half-Cs surface (with extra Cs atoms) and Sb surface, together with the atomic-resolution topographies, are presented in SI 14.

The observation of distinct vortex-core states on different surfaces of $CsV_3Sb_5$ is not trivial. One possible explanation for the robust, finite-distance splitting ZBCP on the 3Q-2×2 Cs surfaces is the presence of MBSs inside the vortex core. In the vortex of a conventional superconductor, discrete energy levels of Caroli-de Gennes-Matricon bound states (CBSs) appear at $E_\mu = \mu\Delta^2/E_F$, with µ the half-integer angular (µ=±1/2, ±3/2, ±5/2, …) momentum number[39]. The spatial distribution of specific CBS scales with its angular momentum number µ, leading to a wave function that peaks at $r_\mu \sim |\mu|/k_F$ away from the vortex center. In most superconductors, the energy spacing of $\mu\Delta^2/E_F$ is too small to be discernible, thus a particle-hole symmetric ZBCP is experimentally observed at the core and it splits when moving right away from the vortex center, resulting in an "X"-type cross[39, 41, 42, 29]. This is different from what we have found here. On the other hand, if Dirac surface states become superconducting, inside a vortex, quasiparticles gain an additional half-odd-integer contribution to the total angular momentum number owing to the intrinsic spin texture of Dirac fermions. Therefore, quasiparticles are integer quantized (µ=0, ±1, ±2…) and MBSs can be regarded as a special zero-energy CBS whose spin degree of freedom is eliminated. Theoretically, the MBS extends in real space with a length scale $\sim v_F/\Delta$ (similar to the superconducting coherence length)[41]. The presence of MBS could change the evolution of the quasiparticle spectra across the vortex, as seen in the cases of $Bi_2Te_3$/$NbSe_2$ heterostructures[29] and $WS_2$ (ref. 44). In $CsV_3Sb_5$, the superconducting gap could open on the bulk bands and topological surface states with tiny electron-doping on the Cs surfaces, and they all contribute to the observed CBSs inside the vortex. The presence of MBSs is expected to enhance the conductance near zero energy within a length scale $\sim v_F/\Delta$ from the vortex center, resulting in the finite-distance ("Y"-type) splitting of the ZBCP. By taken the slope of the calculated surface states of $CsV_3Sb_5$ (~0.18 eV Å; ref. 20) and experimentally extracted superconducting gap (~0.4meV), if exist, the MBSs would extend about 45nm in real space, a value in good agreement with our experimental observation. Compared with iron-based superconductors Fe(Se, Te) (ref. 45, 46) and (Li, Fe)OHFeSe (ref. 47), the relatively large Fermi energies for both bulk bands and Dirac surface band in $CsV_3Sb_5$ impede the observation of discrete CBSs, making the case more similar to that of $Bi_2Te_3$/$NbSe_2$ heterostructure.

Furthermore, the reported edge supercurrent in the Nb/$K_{1-x}V_3Sb_5$ devices[28] and chiral anisotropy in the CDW states[23] indicate that the superconducting state in this family of compounds may also simultaneously break the time-reversal-symmetry (TRS), making the situation more complex. A spin-triplet superconducting state with breaking TRS could also host MBSs inside the vortex core. To further reveal the superconductivity, possible future plans would include Bogoliubov quasiparticle interference measurements, to probe the superconducting gap functions $\Delta_i(k)$, and searching for possible edge states in the *dI/dV* maps. Finally, we would like to emphasize that, although our STM data are consistent with the presence of MBSs inside the vortex core of $CsV_3Sb_5$, further investigations at lower temperature and on doped samples are necessary and more evidence is needed. Moreover, the distinct vortex-core states on the Cs and Sb surfaces, which may depend on local carrier doping[48] or different CDW modulations, would be an interesting topic for future theoretical study.

In summary, we have demonstrated a three-dimensional 2×2×2 CDW state with an energy gap opening near the Fermi level in kagome superconductor $CsV_3Sb_5$. In the superconducting state, we find a ZBCP inside the vortex core that splits off at a large distance away from the vortex center on the Cs surfaces. This unusual splitting behavior, reminiscent of that observed in $Bi_2Te_3/NbSe_2$, could be possibly due to the MBSs inside the vortex core. Our findings may have provided the first glimpse into possible Majorana modes in a kagome superconductor.

**Acknowledgements**

This work is supported by the National Key Research and Development Program of the Ministry of Science and Technology of China (No. 2017YFA0303001, 2017YFA0302904 and 2018YFA0305602), the National Natural Science Foundation of China (Grants No. 11888101, 12074002 and 11804379), the Strategic Priority Research Program of Chinese Academy of Sciences (Grant No. XDB25000000), Anhui Initiative in Quantum Information Technologies (Grant No. AHY160000), the Key Research Program of Frontier Sciences, CAS, China (QYZDYSSW-SLH021). Z.Y.W. is supported by National Natural Science Foundation of China (No. 12074364), the Fundamental Research Funds for the Central Universities (WK3510000012) and USTC start-up fund. L. S. is supported by the Recruitment Program for Leading Talent Team of Anhui Province (2019-16).

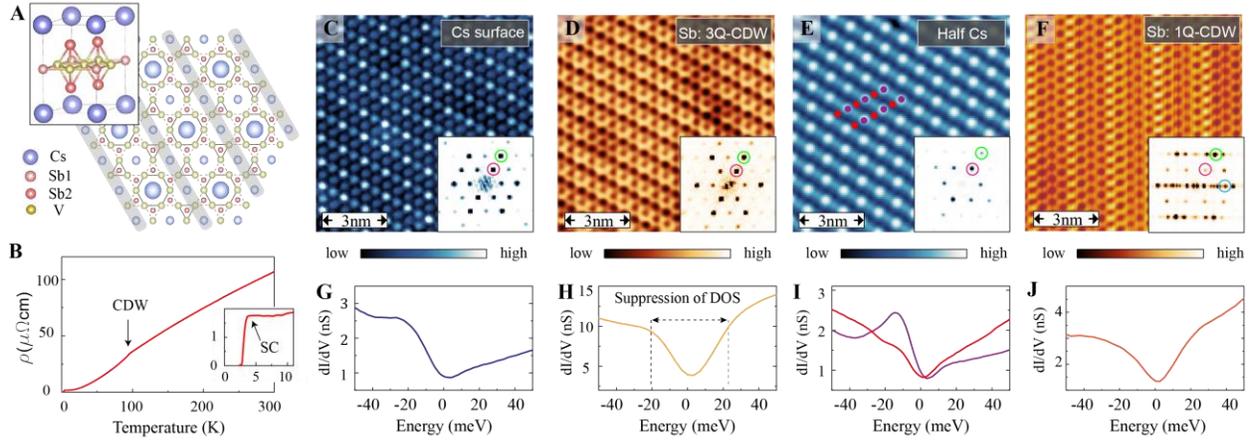

**Figure 1.** (A) Crystal structure of CsV$_3$Sb$_5$ and the illustration of a top view of the lattice. The size of Cs atoms represents the CDW modulation. (B) Temperature-dependent resistivity of CsV$_3$Sb$_5$. A clear anomaly occurs around 94K, indicating the formation of charge order. The superconducting transition occurred at low temperature is shown in the inset. (C-E) Topographies of the Cs, Sb and half-Cs surfaces, respectively. All of them show clear 3Q 2×2 CDW patterns. (F) Topography obtained on Sb surfaces, showing additional 1Q modulation. This 1Q modulation is observed on large Sb surfaces with very few Cs atoms on the top, and it becomes weaker with more surrounding Cs atoms (SI 4). The insets display the Fourier transforms of these topographic images. The lattice peak is marked by green circles, the 3Q-CDW vector in red circles while the 1Q-CDW vector in blue. (G-J) Typical d*I*/d*V* spectra for these surfaces. The purple and pink dots in E denote the locations where spectra are obtained for half-Cs surface. STM setup condition: (C) $V_S$ =-70 mV, $I_t$=80 pA; (D) $V_S$ =-70 mV, $I_t$=120 pA; (E) $V_S$ =-50 mV, $I_t$=700 pA; (F) $V_S$ =-100 mV, $I_t$=150 pA; (G) $V_S$ =-60 mV, $I_t$=150 pA, $V_m$=3 meV; (H) $V_S$ =-50 mV, $I_t$=450 pA, $V_m$=4 meV; (I) $V_S$ =-60 mV, $I_t$=250 pA, $V_m$=3 meV; (J) $V_S$ =-50 mV, $I_t$=130 pA, $V_m$=2 meV.

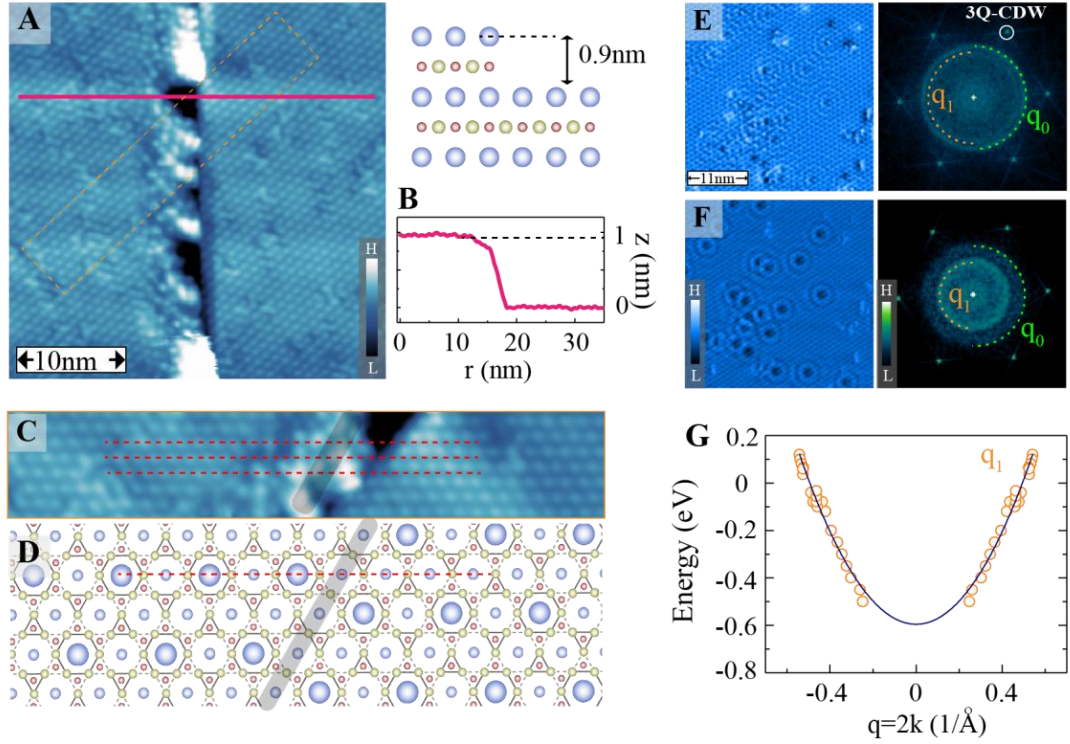

**Figure 2.** (A) Topographic image of the half-Cs surface showing a single-unit-cell step. (B) Height profile obtained along the pink line shown in A. (C) Close-up of the step edge. The dashed pink lines track the chains with CDW modulation on the upper side. A π-phase jump can be observed between the upper and lower sides. (D) Illustration of the CDW patterns near a single-unit-cell step. (E-F) Normalized *dI/dV* maps (*dI/dV* (*r*, eV)/(I(*r*, eV)/V))and their FFTs at -80mV and -300mV, respectively. The white circle denotes one of the 3Q-CDW wave vectors. The clearly-dispersing vector is marked as $q_1$. (G) The energy dispersion of $q_1$ extracted from line-cuts of the FFTs (SI 7) of and a parabolic fit, $E = 9.77 \times k^2 - 0.6$. STM setup condition: (A) $V_S$ =-100 mV, $I_t$=80 pA; (E, F) $V_S$ =-300 mV, $I_t$=450 pA, $V_m$ =4 mV.

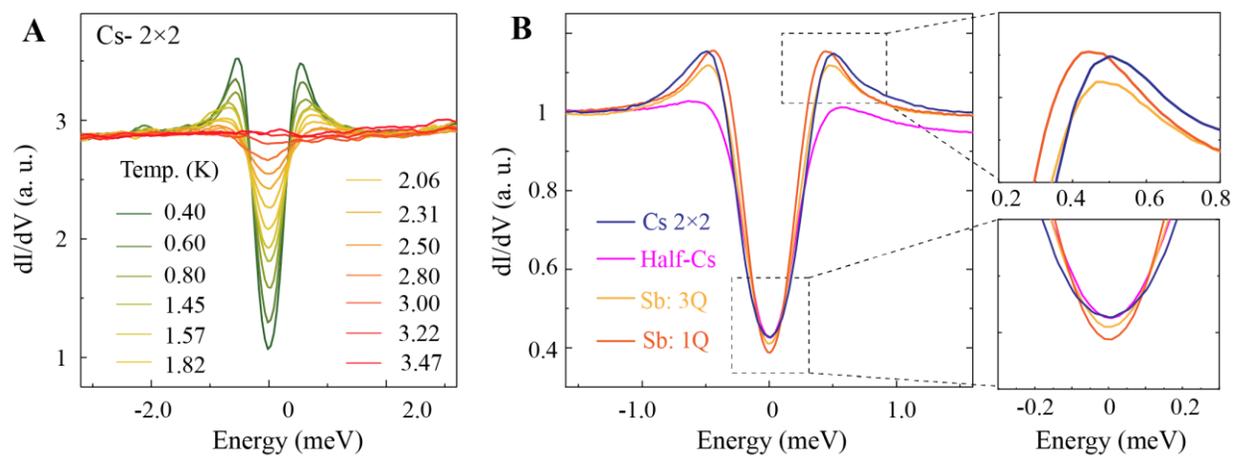

**Figure 3.** (A) Temperature evolution of the superconducting gap. (B) Superconducting spectra obtained on different types of surfaces. The spectra of half-Cs, Sb 3Q-2×2 and Sb 1Q regions are obtained in one field of view with the same tip. STM setup condition: (A, B) $V_s = -2$ mV, $I_t = 200$ pA, $V_m = 0.1$ mV.

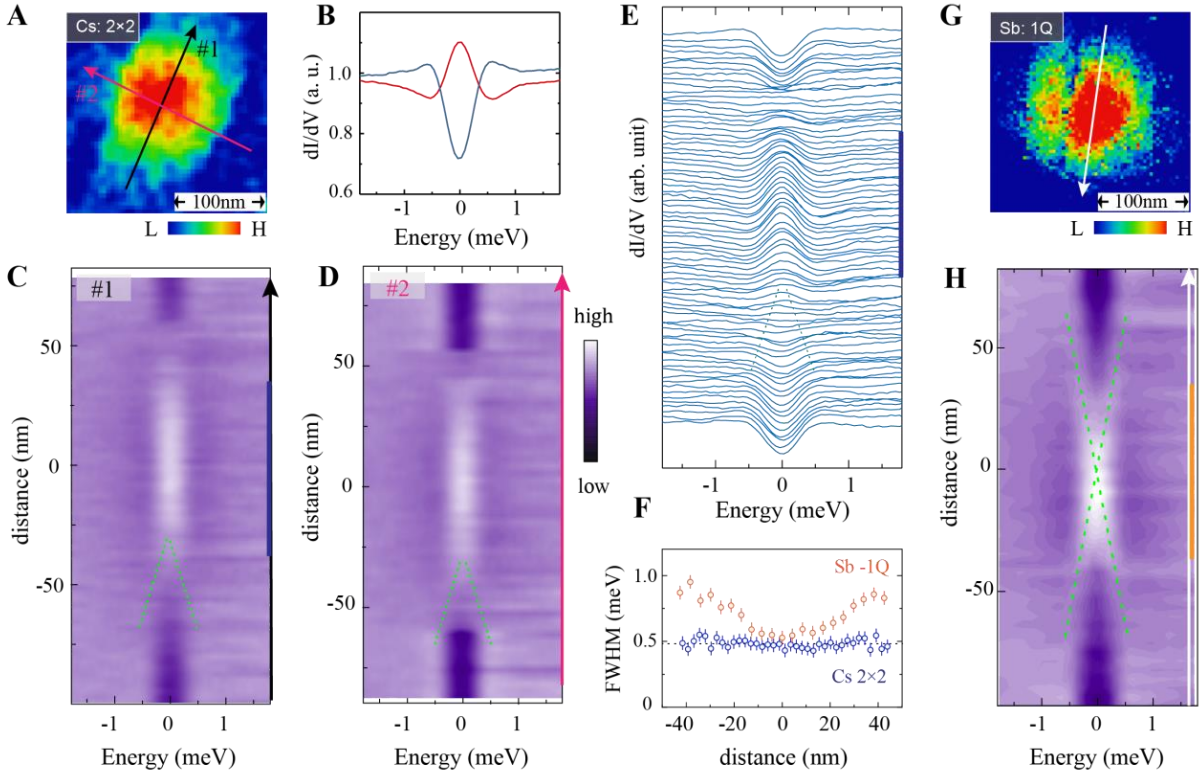

**Figure 4.** (A) dI/dV map showing a superconducting vortex on the Cs surface with 3Q-2×2 CDW. (B) Tunneling spectra obtained in the vortex core (red) and outside the vortex (dark-blue). (C, D) Color map of the spatially resolved dI/dV spectra across the vortex. The traces are marked by the dark and pink arrows as shown in A. (E) The waterfall-plot of C, vertically offset for clarity. (F) Spatial dependence of FWHM of the ZBCP when moving away from the vortex core for Cs surface (blue) and Sb surface (yellow). The error bar denotes the energy step used in the measurement. (G) dI/dV map showing a vortex on the Sb surface with additional 1Q- CDW. (H) Spatial evolution of the dI/dV spectra across the vortex. Dashed lines in light green track the start points of the peak splitting. STM setup condition: (A) $V_S$ =-5 mV, $I_t$=200 pA, $V_m$ =0.5 mV; (G) $V_S$ =-6 mV, $I_t$=200 pA, $V_m$ =0.5 mV; (C, D, H) $V_S$ =-5 mV, $I_t$=200 pA, $V_m$ =0.1 mV.

# Supplementary information for
# "Three-dimensional charge density wave and robust zero-bias-peak inside the superconducting vortex core of a kagome superconductor $CsV_3Sb_5$"

Zuowei Liang, Xingyuan Hou, Fan Zhang, Wanru Ma, Ping Wu, Zongyuan Zhang, Fanghang Yu, J.-J. Ying, Kun Jiang, Lei Shan, Zhenyu Wang and X. -H. Chen

**This PDF file includes:**

1. **Sample characterizations by XRD**

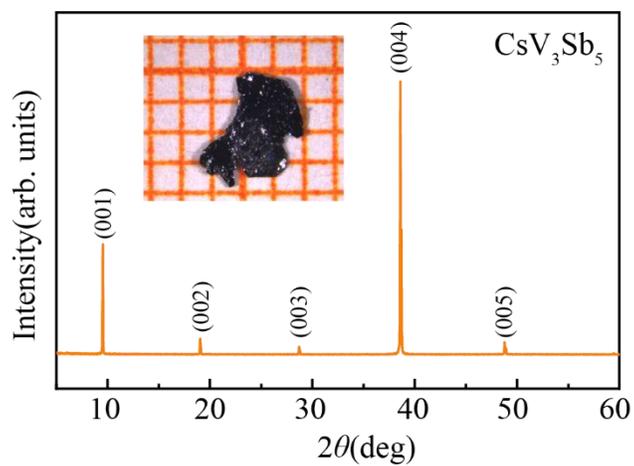

**Figure S1.** X-ray diffraction data of $CsV_3Sb_5$ single crystals with the corresponding Miller indices (00L) in parentheses. The inset shows a photo of one single crystal, and the grid spacing of the background is 1mm.

## 2. A joint area among Sb surface, half-Cs surface and Cs surface

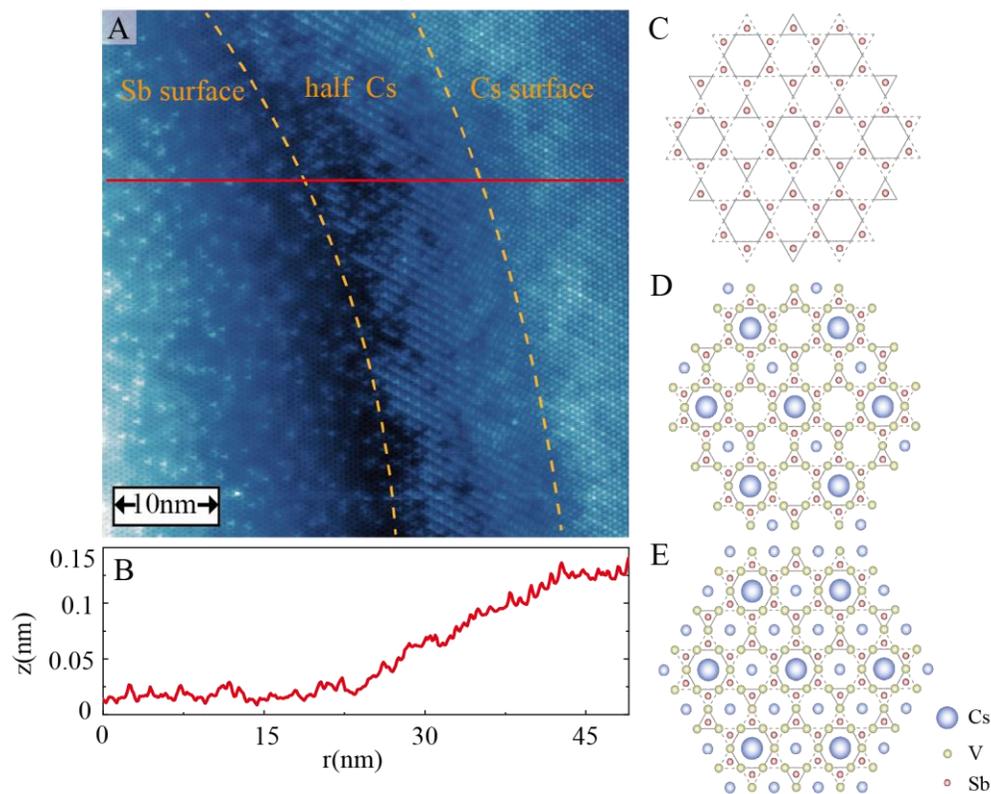

**Figure S2.** A) STM topography of a joint area among Sb, half-Cs and Cs surfaces. Even for a Cs clustered-chain with only several Cs atoms floating on the Sb surface, it still exhibits clear charge modulation. B) Height profile along the red line shown in A. C)-E) Illustrations of Sb surface, half-Cs surface and Cs surface, respectively. The size of the Cs atoms indicates the 2×2 modulation pattern, and the solid-lines indicate the formation of inverse star of David in the underlying V plane.

## 3. Bias-dependent topographic images of half-Cs surface

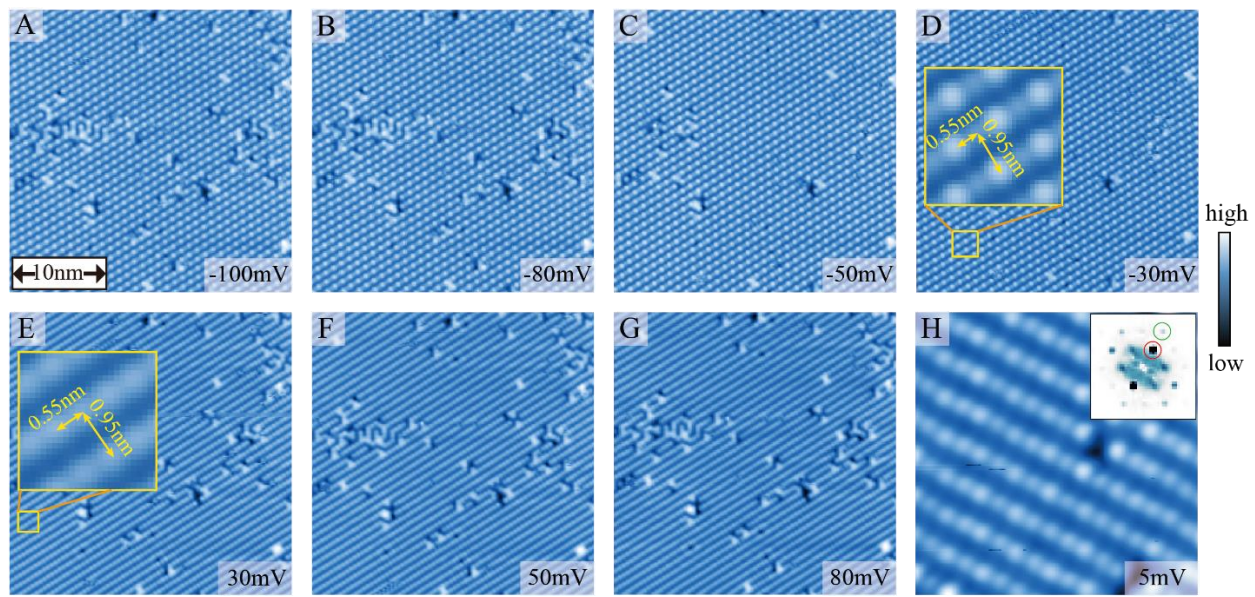

**Figure S3.** A)-H) STM topographies obtained on half-Cs surface at various sample bias voltages. At higher negative biases, the dark Cs atoms in the chains are invisible, leaving only bright Cs atoms in the images (This is the case of Fig. 2A). At high positive biases, all the atoms in the chains can be seen and the modulation is weak. The insets in D and E show the zoomed-in views of yellow squares. The inset of H displays FT that shows the CDW modulation at +5mV (red circle).

## 4. 1Q-CDW modulations observed on Sb surface

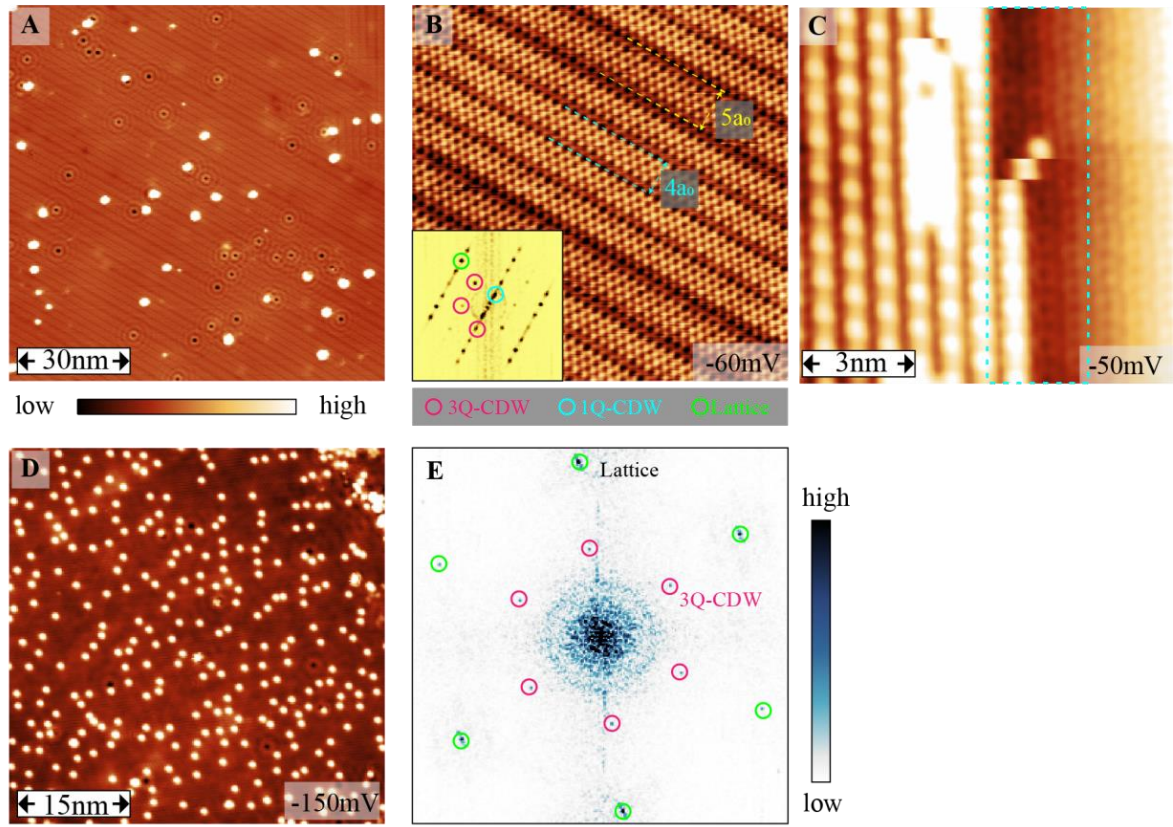

**Figure S4.** A) Large STM topography of Sb surface with unidirectional (1Q) modulation. B) Atomic resolved topography in a 10nm×10nm field of view. The inset shows its FFT. Both the 3Q- and 1Q-modulation peaks can be observed. We note that the 1Q modulation is observed on large Sb surfaces with very rare Cs atoms on the top. Near the boundary of half-Cs and Sb surfaces, the 3Q-2×2 modulation becomes more visible (marked with dashed box in Fig. S4C). With more Cs atoms on the top, the 1Q feature becomes much weaker and even invisible, and in the FFT one can only see 3Q-CDW peaks at half magnitude of the Bragg peak (Fig. S4D and E).

## 5. Line-cuts of spectra and atomic-resolved spectroscopic images

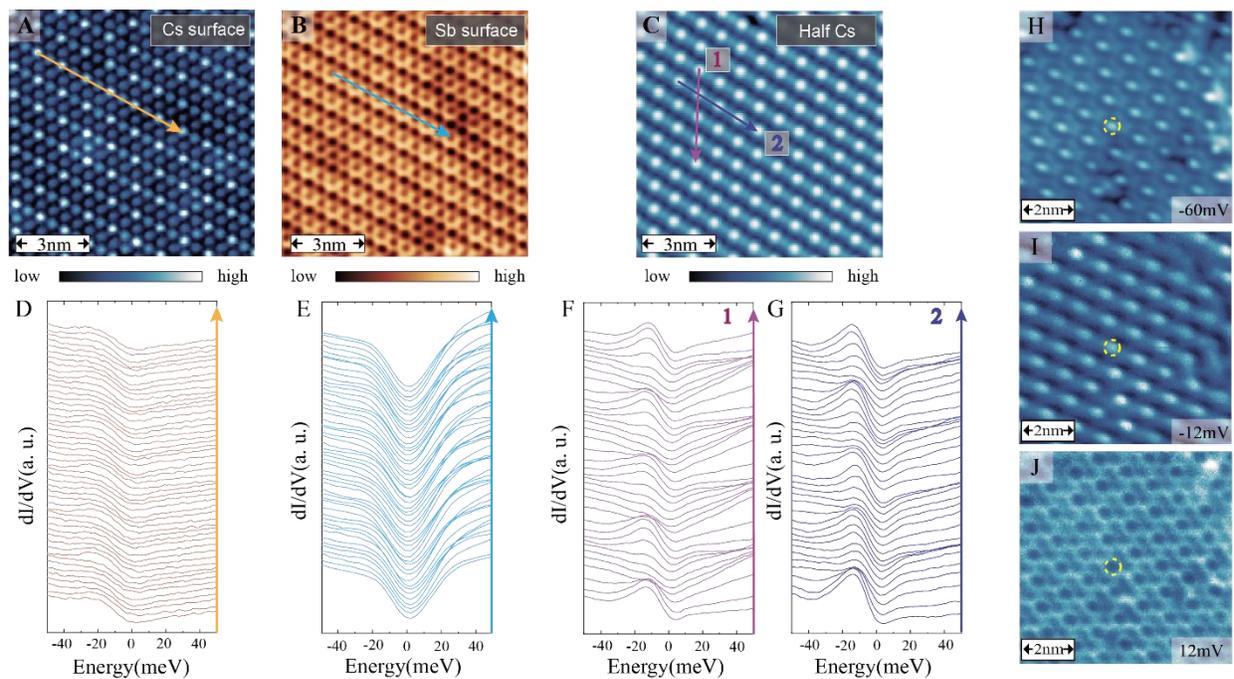

**Figure S5.** A-C) Topographies of Cs, Sb and half-Cs surfaces. D-G) linecuts of *dI/dV* spectra measured along the lines shown in A-C. H) STM topography of half-Cs surface. I, J) The simultaneously obtained *dI/dV* maps at -12mV and 12mV in the same field of view of H. A clear charge modulation, which reverses its contrast can be clearly seen in these maps.

## 6. Temperature-dependence of the tunneling spectra

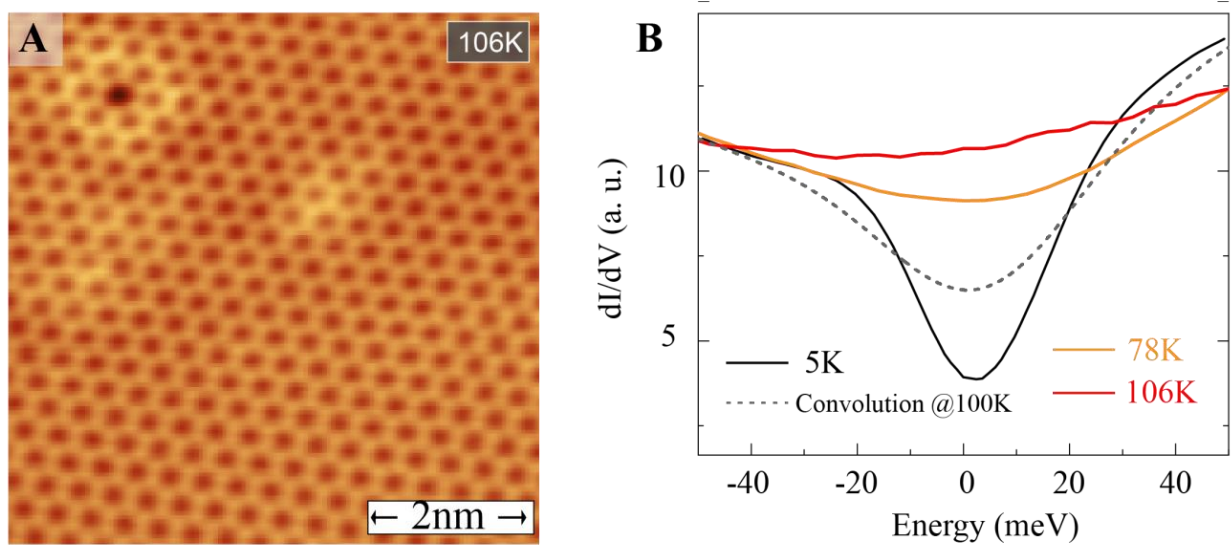

**Figure S6.** A) STM topography of Sb surface at 106K. B) The temperature-dependent *dI/dV* spectra obtained on the Sb surfaces. At 106K ($T_{CDW}$~94K), the DOS is almost flat near the Fermi energy, and we cannot see a clear gap. The dashed curve denotes the convoluted 5K spectra by Fermi-Dirac distribution function (at 100K). It is clear that the convoluted data at 100K still exhibit a suppression of DOS, indicating that the measured featureless DOS at 106 cannot be explained by thermal broadening.

## 7. Displacement analysis near a single-unit-cell step

Fourier transform of a topographic image with a prefect lattice shows sharp peaks in the reciprocal space. When two parts that have a relatively shift are included in the field of view, the sharp peaks split into two peaks [1, 2]. A displacement analysis allows us to extract the related phase shift between these two parts [2]. The algorithm we use here was first demonstrated by Hÿch, Snoeck, and Kilaas to measure displacement and strain in high-resolution electron microscope (HREM) images [3], and then adapted by the STM community where it is known as the "Lawler-Fujita algorithm" [4].

In the FFT of Fig. 2A, the peak locations (corresponding to the bright Cs atoms, for simplicity here we call them CDW peaks, though in the strict sense "CDW" is only valid in one direction due to the missing atoms) of upper and lower layers can express as $\vec{g}, \vec{g}'$, respectively. The relationship between them can be described as $\vec{g}' = \vec{g} + \delta\vec{g}(\vec{r})$. When looking at Fourier component of the lattice, we can find that it takes the form of $I_{lattice'(\vec{r})} = A_{g_A}(\vec{r})e^{2\pi i \vec{g}_A \cdot \vec{r}} e^{2\pi i \delta \vec{g}_A(\vec{r}) \cdot \vec{r}} + A_{g_B}(\vec{r})e^{2\pi i \vec{g}_B \cdot \vec{r}} e^{2\pi i \delta \vec{g}_B(\vec{r}) \cdot \vec{r}} + A_{g_C}(\vec{r})e^{2\pi i \vec{g}_C \cdot \vec{r}} e^{2\pi i \delta \vec{g}_C(\vec{r}) \cdot \vec{r}}$. If choosing a pair of CDW points, we can get the modulation information of one direction $I_{lattice'_i} = A_{g_i}(\vec{r})e^{2\pi i \vec{g}_i \cdot \vec{r}} e^{2\pi i \delta \vec{g}_i(\vec{r}) \cdot \vec{r}}; i = A, B, C$. By multiplying a reference signal $e^{-2\pi i \vec{g}_i \cdot \vec{r}}; i = A, B, C$, The periodicity of the lattice can be removed, and we get the $I_{shift_i}(\vec{r}) = A_{g_i}(\vec{r})e^{2\pi i g_i \delta(\vec{r}) \cdot \vec{r}}; i = A, B, C$. Then by taking the phase of the remaining image, $\vec{g}_i \cdot \delta\vec{r}$ can be isolated up to $\pm\pi$, $\text{Phase}[I_{shift_i}(\vec{r})] = 2\pi\delta g_i(\vec{r}) \cdot \vec{r} \mod \pi; i = A, B, C$. If lower layer has a half CDW "unit-cell" offset from upper layer along the C direction, we can get $\delta\vec{g}_C(\vec{r}) \cdot \vec{r} = 0$, $\delta\vec{g}_A(\vec{r}) \cdot \vec{r} = 1/2$, $\delta\vec{g}_B(\vec{r}) \cdot \vec{r} = 1/2$. In A and B direction, the phase difference between upper and lower layers will be $\pi$. Along the C direction, there will be no phase difference.

We apply this procedure to the topography shown in Fig. 2A, and plot the result in Fig. S7A-C. Two of the peaks (responding to the bright Cs atoms in the CDW pattern) split, as shown in Fig. S7A. A linecut of the phase map (Fig.S7B) clearly exhibits a phase different of $\pi$ across the step. Performing the same procedure on the other splitting peak, we also a $\pi$ phase jump.

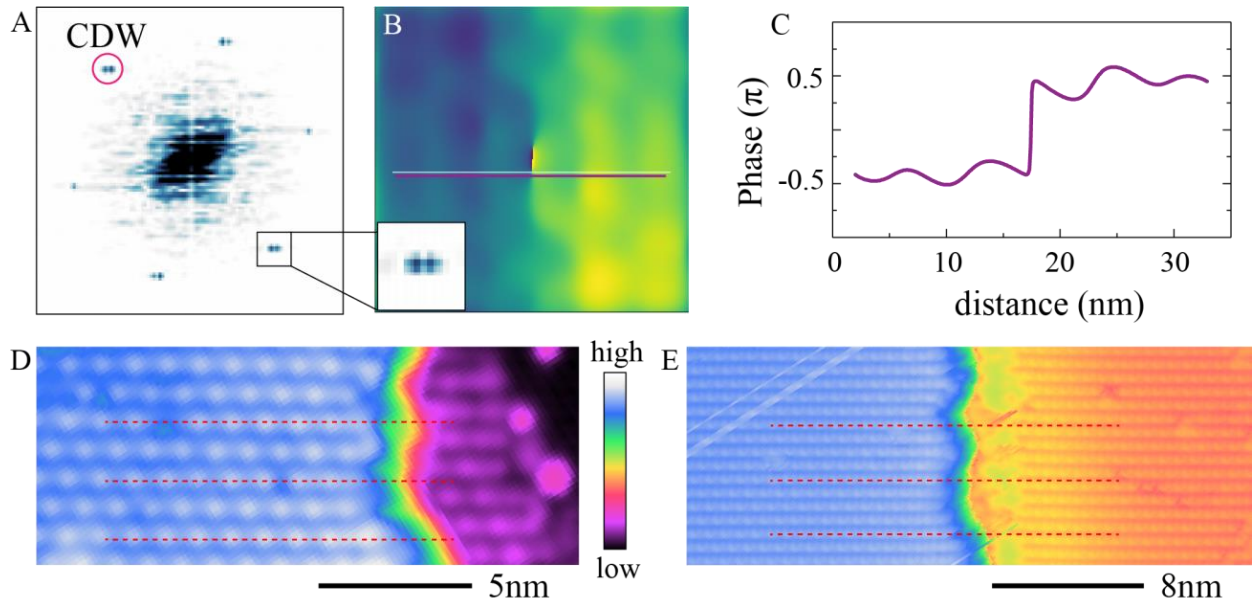

**Figure S7.** A) The FFT of Fig. 2A showing the splitting of the CDW peaks (in the strict sense "CDW" is only valid in one direction due to the missing atoms), indicating a relative phase shift of the CDW pattern. B) Phase map from one of the split CDW peaks (inset). A general π-phase shift across the single-unit-cell step edge can be seen. C) Line-cut profiles in the phase map. D) and E) More single-unit-cell step measurements. Guided by the red dash lines, it is clear that there is a phase shift across the step in both of the images.

## 8. More QPI data of half-Cs surface

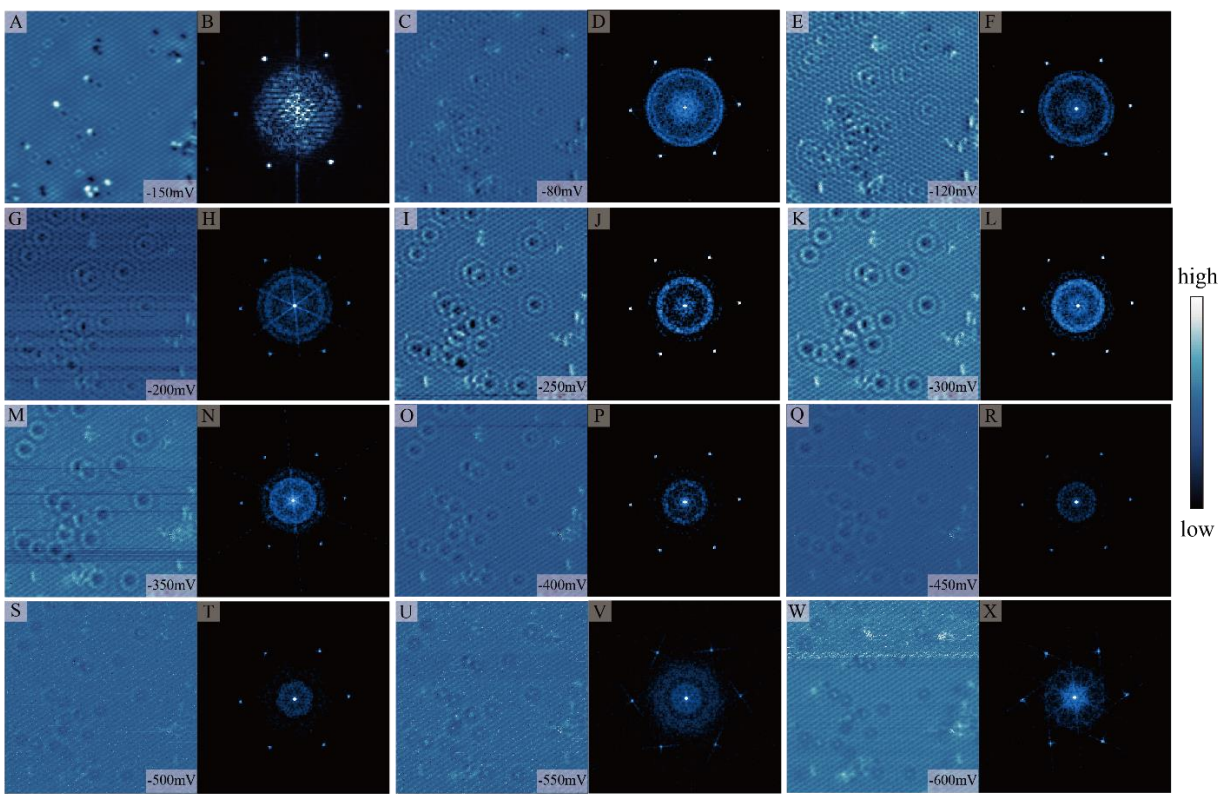

**Figure S8.** A)-B) STM topography of half-Cs surface and its FFT image. C)-X) *dI/dV* maps and their FFTs after symmetrization in the field of view of A).

## 9. Fittings of tunneling spectra with different gap functions

Only one main pair of well-resolved coherence peaks could be recognized on the spectrum measured at 0.4 K. Hence, we firstly carry out the Dynes model [5] with an isotropic s-wave or an anisotropic s-wave gap function to fit the spectrum which obeys:

$$dI/dV = \frac{1}{2\pi}\int d\varepsilon \int_0^{2\pi} d\theta \frac{df(\varepsilon)}{d\varepsilon}|_{\varepsilon+eV} abs(Re[\frac{\varepsilon-i\Gamma}{(\varepsilon-i\Gamma)^2-\Delta(\theta)^2}]),$$

where $f(\varepsilon)$ is the Fermi distribution function, $\Delta(\theta)$ is the hypothetical gap function and $\Gamma$ is the scattering rate. In A), we show the shape of the hypothetical s-wave gap with a constant size of 0.32 meV.

Seeking to a better fitting result, more hypothetical gap functions are considered. Considering the lattice symmetry of CsV$_3$Sb$_5$, we then try an anisotropic s-wave gap function with a six-fold symmetry in B): $\Delta(\theta) = \Delta_0[\alpha|\cos(3\theta)| + (1-\alpha)]$. The gap anisotropy is determined by $\alpha$. The maximum value of the gap $\Delta_0$ we adopt here is 0.41 meV and the value of $\alpha$ is 0.6.

To contain more information, we also employ an extended two-band Dynes model according to the fermi surface topology of CsV$_3$Sb$_5$ [6]: $dI/dV = w_S dI_S/dV + (1-w_S) dI_L/dV$, where $\Delta_S$ is the smaller superconducting gap with the scattering rate $\Gamma_S$, $w_S$ is the related spectral weight. The two isotropic gaps we obtain in C) are: $\Delta_S$= 0.3 meV, $\Delta_L$= 0.36 meV, $and$ $w_S$ = 0.35.

For each case, we minimize the deviation between the experimental data and the fitting curves, and calculate their difference (shown in the lower panels of Fig. S9 D-F. When focusing on the low energy gap feature, we find that the inclusion of anisotropy in the superconducting gap function could clearly improve the fitting. Therefore, the fits to the superconducting spectrum seem most consistent with an anisotropic superconducting order parameter for CsV$_3$Sb$_5$, though we cannot make any conclusion on the existence of node from the existing data.

Figure S9 G)-I) show the temperature dependence of the normalized dI/dV spectra. The data are well consistent with the calculations based on the extended Denes model mentioned above (denoted by the solid red lines). The temperature dependent gaps extracted from the fitting are exhibited in J)-L). One can see that they could be well-described by the BCS theory. It is worth mentioning that these particular gap functions are selected just for taking into account the possible anisotropy (even node) or multi-band effect.

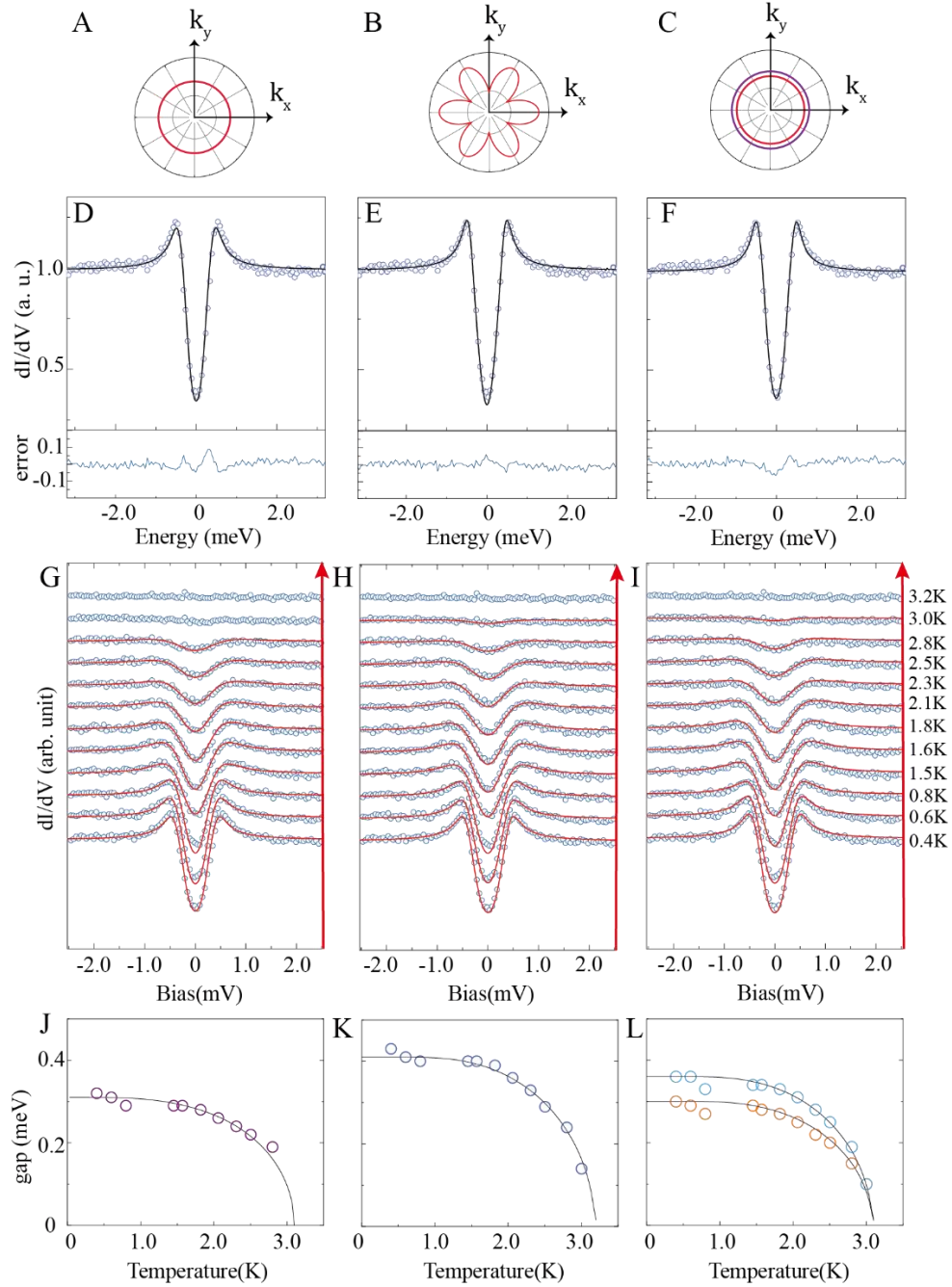

**Figure S9.** A)-C) The assumed gap functions, namely, isotropic, anisotropic and two isotropic s-wave gaps in the fittings. D)-F) Fitting results with the Dynes model based on three different superconducting gap functions, to the data obtained at 0.4K. The low panels show the deviations between the experimental data and the fitting results. G-I) Temperature dependence of the normalized dI/dV spectra (cadet blue circles) and the fitting results with the Dynes model (solid red lines) for isotropic, anisotropic and two isotropic gaps, respectively. J)-L) Temperature dependence of the gap values extracted from the fitting with different gap functions. The solid lines denote the prediction of the superconducting gap at different temperature based on the BCS model.

## 10. A direct comparison of vortex core states among NbSe$_2$, CsV$_3$Sb$_5$ and Bi$_2$Te$_3$/NbSe$_2$

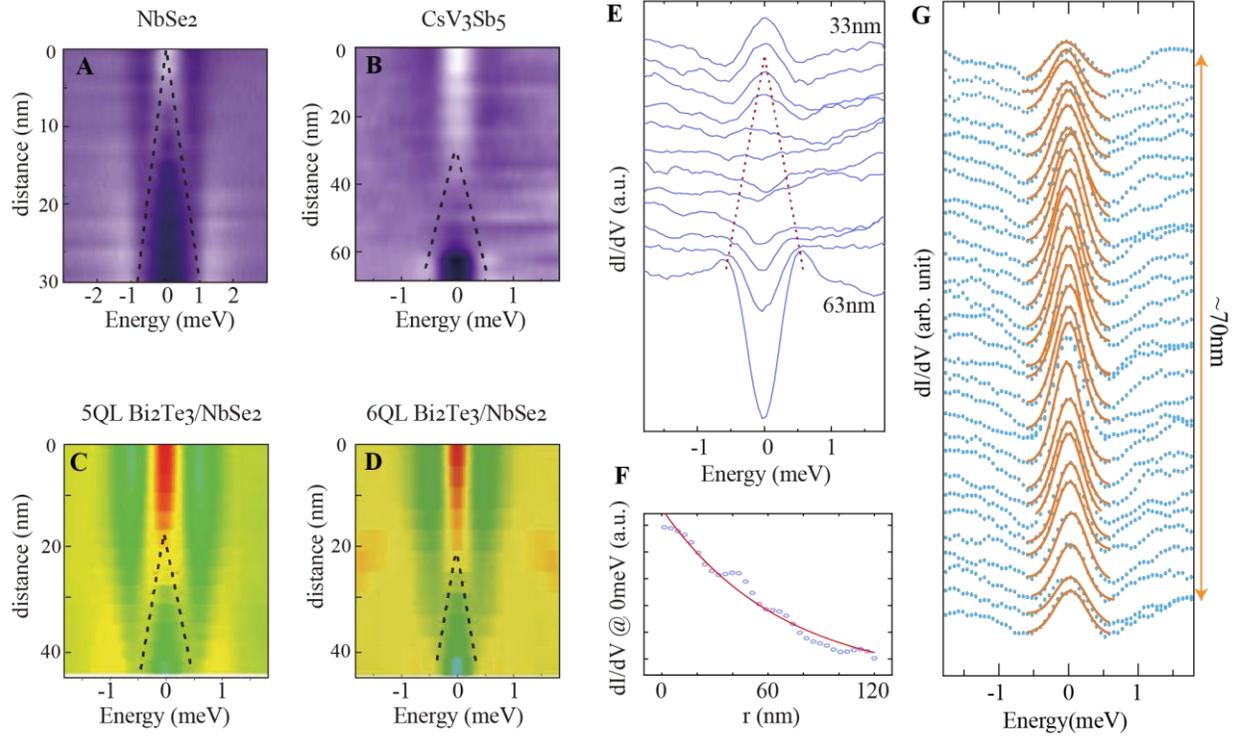

**Figure S10.** A-D) Color maps of the spatially resolved *dI/dV* spectra across the vortex for NbSe$_2$, CsV$_3$Sb$_5$, 5QL- and 6QL- Bi$_2$Te$_3$/NbSe$_2$ heterostructures, respectively. "X"-type peak splitting can be found for NbSe$_2$, while in others one find "Y"-type splitting. Data shown in C and D are from [7]. E) Zoom-in of the splitting spectra of CsV$_3$Sb$_5$, 33nm-63nm from the vortex core center. F, Spatial size of the zero-bias-peak of CsV$_3$Sb$_5$. An exponential decay fitting yields a value of ξ=67.8nm. G) The Gaussian fits of ZBCP for r<35nm.

## 11. Energy-resolution calibration of the instrument

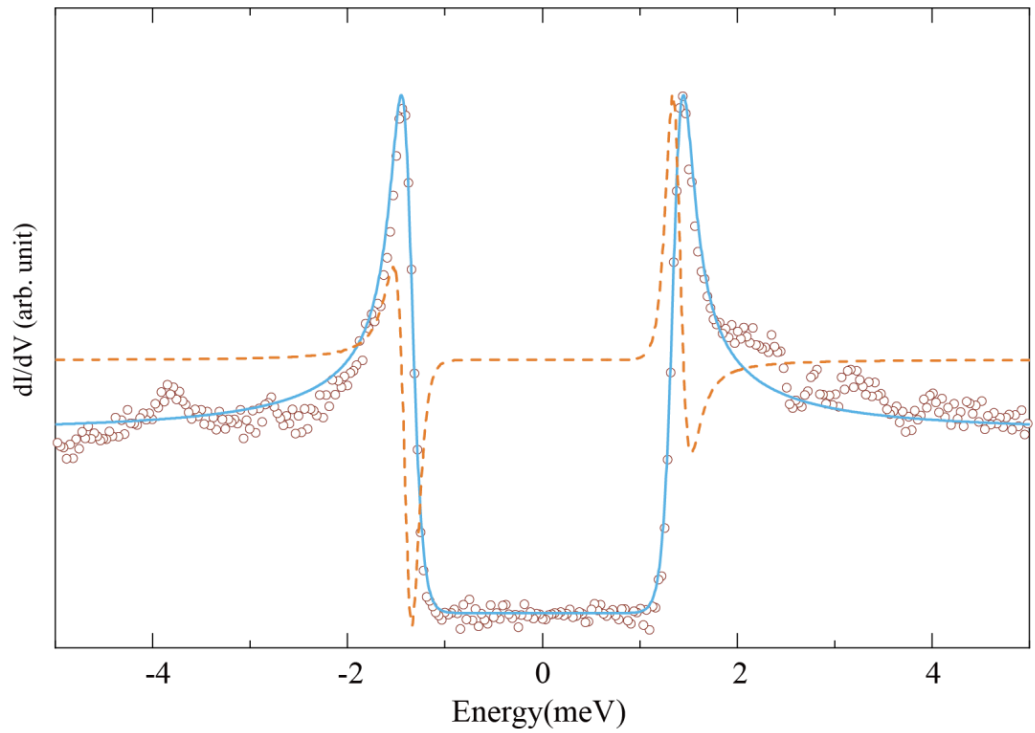

**Figure S11.** A) The superconducting spectrum obtained on a Pb surface at 0.4K with our STM. The blue curve denotes a BCS fitting. Fitting parameters: Δ=1.38 meV, Γ=0meV, $T_{eff} = 0.66$ K. The total broadening of the superconducting gap edge is about 0.23meV.

## 12. Clear superconducting gap at zero field along line 1 of Fig. 4A

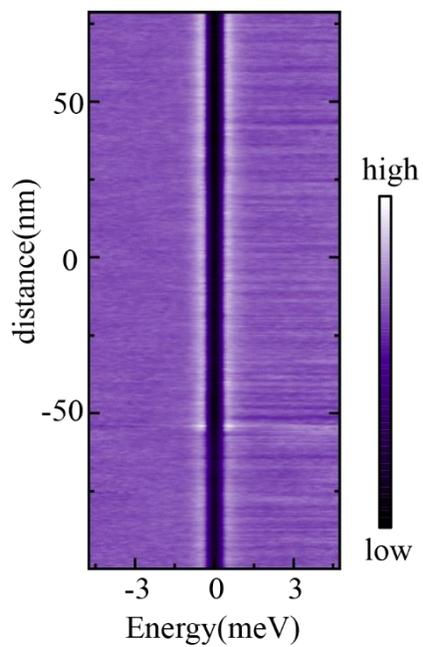

**Figure S12.** A clean superconducting gap can be observed at zero field along the same trace of line1(#1) in Fig.4A.

## 13. More vortex data taken on the Cs surfaces

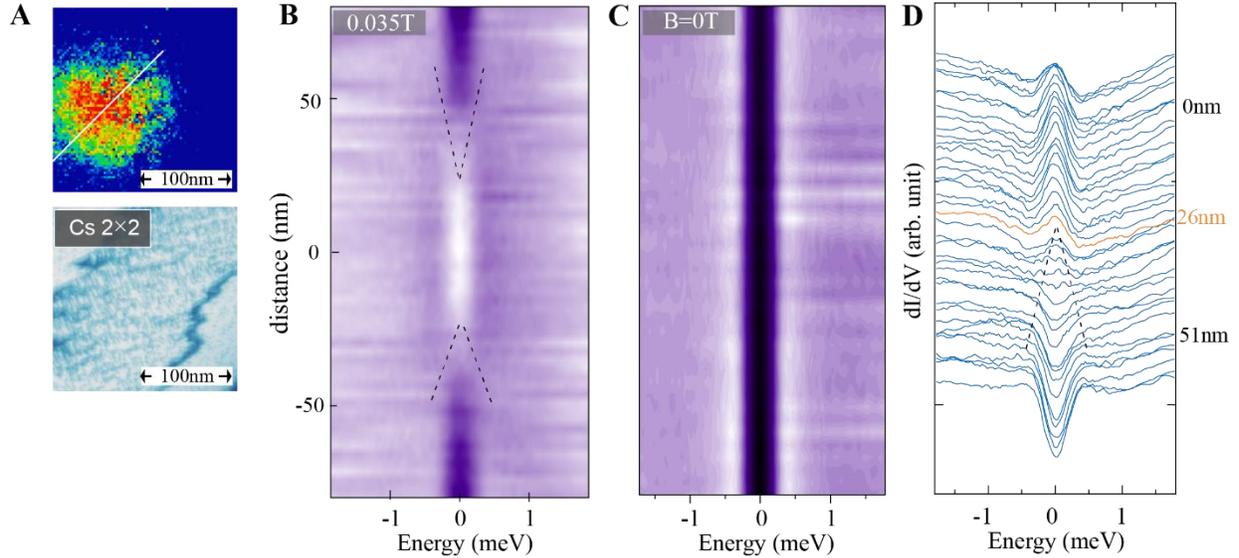

**Figure S13.** A) Another superconducting vortex obtained on the Cs surface with 3Q-2×2 CDW, in a different sample with different tip, with B=0.035T. B) Spatial evolution of the dI/dV spectra across the vortex, showing similar finite-distance splitting of the ZBCP (Close-up is shown in D). C) spectra taken at zero field along the same trace of B. D) Zoom-in of the splitting spectra, and the start point of splitting is located ~26 nm away from the vortex center. Finite-distance splitting of ZBCP is repeatable once the vortex is located on the Cs surfaces.

## 14. Evolution of the vortex-core state on different surfaces

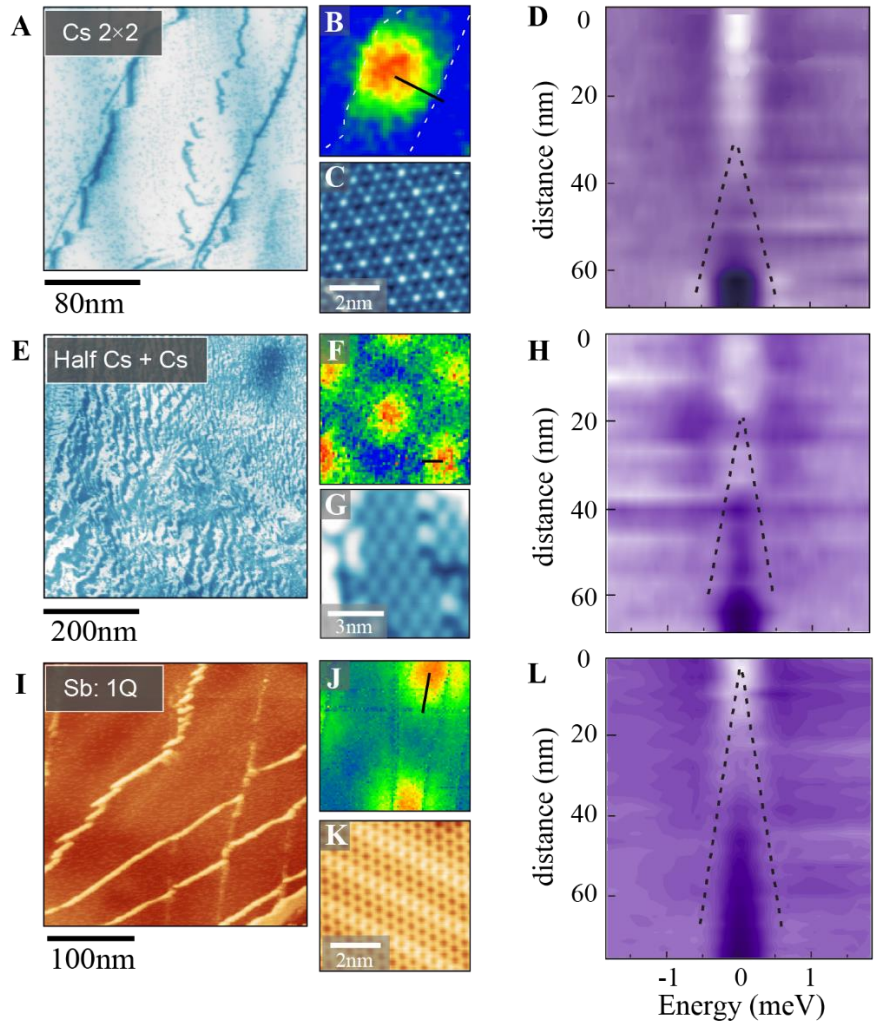

**Figure S14.** A) Topographic image of a Cs surface with 3Q-2×2 CDW where the vortex shown in B has been imaged. The dashed white lines in B denote the crack-like features due to the missing Cs atoms. C) Atomic resolved topography showing the 2×2 CDW pattern. D) The color image of spectra measured along the dark line in B from the vortex core center. E-H), similar plots as A-D, for the half-Cs surface with extra Cs atoms (bright spots). I-L) Similar plots for the Sb surfaces. Dashed dark lines guide the start point of the splitting.

The largely-explored Sb surfaces usually exhibit additional 1Q modulation, which is in contrast to the Cs surfaces that show only 3Q-2×2 CDW. As one can see, the start point of the splitting locates far away from the core center on the Cs surfaces, while near the center on the Sb surfaces. On the half-Cs surfaces, the spectra are relatively more disordered, probably due to a large scattering as suggested from the superconducting spectra (Fig.3B). The distinct vortex-core states on the Cs and Sb surfaces may depend on local carrier doping of Cs or the effect of different CDW modulations, which could be an interesting topic for future theoretical study.